\documentclass[longauth]{aa} % for the long lists of affiliations 
\usepackage{natbib}
\bibpunct{(}{)}{;}{a}{}{,} % to follow the A&A style
\usepackage{graphicx}
\usepackage{txfonts}
\usepackage{textcomp} % Oma lisäys astemerkkiä (textdegree) varten.

\graphicspath{ {./Images/} }

\begin{document}

   \title{\textit{Euclid}: Identification of asteroid streaks in\\simulated images using \texttt{StreakDet} software\thanks{This paper is published on behalf of the Euclid Consortium.}}

%% please do not edit the author list -- contact ECEB Bureau for changes
\author{M.~Pöntinen$^{1}$\thanks{\email{mikko.pontinen@helsinki.fi}}, M.~Granvik$^{1,2}$, A.A.~Nucita$^{3,4}$, L.~Conversi$^{5,6}$, B.~Altieri$^{6}$, N.~Auricchio$^{7}$, C.~Bodendorf$^{8}$, D.~Bonino$^{9}$, M.~Brescia$^{10}$, V.~Capobianco$^{9}$, J.~Carretero$^{11}$, B.~Carry$^{12}$, M.~Castellano$^{13}$, R.~Cledassou$^{14}$, G.~Congedo$^{15}$, L.~Corcione$^{9}$, M.~Cropper$^{16}$, S.~Dusini$^{17}$, M.~Frailis$^{18}$, E.~Franceschi$^{7}$, M.~Fumana$^{19}$, B.~Garilli$^{19}$, F.~Grupp$^{8}$, F.~Hormuth$^{20}$, H.~Israel$^{21}$, K.~Jahnke$^{22}$, S.~Kermiche$^{23}$, T.~Kitching$^{16}$, R.~Kohley$^{6}$, B.~Kubik$^{24}$, M.~Kunz$^{25}$, R.~Laureijs$^{26}$, P.~B.~Lilje$^{27}$, I.~Lloro$^{28,29}$, E.~Maiorano$^{7}$, O.~Marggraf$^{30}$, R.~Massey$^{31}$, M.~Meneghetti$^{7}$, G.~Meylan$^{32}$, L.~Moscardini$^{7,33,34}$, C.~Padilla$^{11}$, S.~Paltani$^{35}$, F.~Pasian$^{18}$, S.~Pires$^{36}$, G.~Polenta$^{37}$, F.~Raison$^{8}$, M.~Roncarelli$^{7,33}$, E.~Rossetti$^{33}$, R.~Saglia$^{8}$, P.~Schneider$^{30}$, A.~Secroun$^{23}$, S.~Serrano$^{28,29}$, G.~Sirri$^{38}$, A.N.~Taylor$^{15}$, I.~Tereno$^{39,40}$, R.~Toledo-Moreo$^{41}$, L.~Valenziano$^{7,38}$, Y.~Wang$^{42}$, M.~Wetzstein$^{8}$, J.~Zoubian$^{23}$}

%% please do not edit the affiliation list -- contact ECEB Bureau for changes
\institute{$^{1}$ Department of Physics, P.O. Box 64, 00014 University of Helsinki, Finland\\
$^{2}$ Asteroid Engineering Laboratory, Onboard Space Systems, Lule\aa{} University of Technology, Box 848, 98128 Kiruna, Sweden\\
$^{3}$ INFN, Sezione di Lecce, Via per Arnesano, CP-193, I-73100, Lecce, Italy\\
$^{4}$ Department of Mathematics and Physics E. De Giorgi, University of Salento, Via per Arnesano, CP-I93, I-73100, Lecce, Italy\\
$^{5}$ European Space Agency/ESRIN, Largo Galileo Galilei 1, 00044 Frascati, Roma, Italy\\
$^{6}$ ESAC/ESA, Camino Bajo del Castillo, s/n., Urb. Villafranca del Castillo, 28692 Villanueva de la Ca\~nada, Madrid, Spain\\
$^{7}$ INAF-Osservatorio di Astrofisica e Scienza dello Spazio di Bologna, Via Piero Gobetti 93/3, I-40129 Bologna, Italy\\
$^{8}$ Max Planck Institute for Extraterrestrial Physics, Giessenbachstr. 1, D-85748 Garching, Germany\\
$^{9}$ INAF-Osservatorio Astrofisico di Torino, Via Osservatorio 20, I-10025 Pino Torinese (TO), Italy\\
$^{10}$ INAF-Osservatorio Astronomico di Capodimonte, Via Moiariello 16, I-80131 Napoli, Italy\\
$^{11}$ Institut de F\'{i}sica d’Altes Energies (IFAE), The Barcelona Institute of Science and Technology, Campus UAB, 08193 Bellaterra (Barcelona), Spain\\
$^{12}$ Universit\'e C\^{o}te d'Azur, Observatoire de la C\^{o}te d'Azur, CNRS, Laboratoire Lagrange, Bd de l'Observatoire, CS 34229, 06304 Nice cedex 4, France\\
$^{13}$ INAF-Osservatorio Astronomico di Roma, Via Frascati 33, I-00078 Monteporzio Catone, Italy\\
$^{14}$ Centre National d'Etudes Spatiales, Toulouse, France\\
$^{15}$ Institute for Astronomy, University of Edinburgh, Royal Observatory, Blackford Hill, Edinburgh EH9 3HJ, UK\\
$^{16}$ Mullard Space Science Laboratory, University College London, Holmbury St Mary, Dorking, Surrey RH5 6NT, UK\\
$^{17}$ INFN-Padova, Via Marzolo 8, I-35131 Padova, Italy\\
$^{18}$ INAF-Osservatorio Astronomico di Trieste, Via G. B. Tiepolo 11, I-34131 Trieste, Italy\\
$^{19}$ INAF-IASF Milano, Via Alfonso Corti 12, I-20133 Milano, Italy\\
$^{20}$ von Hoerner \& Sulger GmbH, Schlo{\ss}Platz 8, D-68723 Schwetzingen, Germany\\
$^{21}$ Universit\"ats-Sternwarte M\"unchen, Fakult\"at f\"ur Physik, Ludwig-Maximilians-Universit\"at M\"unchen, Scheinerstrasse 1, 81679 M\"unchen, Germany\\
$^{22}$ Max-Planck-Institut f\"ur Astronomie, K\"onigstuhl 17, D-69117 Heidelberg, Germany\\
$^{23}$ Aix-Marseille Univ, CNRS/IN2P3, CPPM, Marseille, France\\
$^{24}$ Univ Lyon, Univ Claude Bernard Lyon 1, CNRS/IN2P3, IP2I Lyon, UMR 5822, F-69622, Villeurbanne, France\\
$^{25}$ Universit\'e de Gen\`eve, D\'epartement de Physique Th\'eorique and Centre for Astroparticle Physics, 24 quai Ernest-Ansermet, CH-1211 Gen\`eve 4, Switzerland\\
$^{26}$ European Space Agency/ESTEC, Keplerlaan 1, 2201 AZ Noordwijk, The Netherlands\\
$^{27}$ Institute of Theoretical Astrophysics, University of Oslo, P.O. Box 1029 Blindern, N-0315 Oslo, Norway\\
$^{28}$ Institute of Space Sciences (ICE, CSIC), Campus UAB, Carrer de Can Magrans, s/n, 08193 Barcelona, Spain\\
$^{29}$ Institut d’Estudis Espacials de Catalunya (IEEC), 08034 Barcelona, Spain\\
$^{30}$ Argelander-Institut f\"ur Astronomie, Universit\"at Bonn, Auf dem H\"ugel 71, 53121 Bonn, Germany\\
$^{31}$ Centre for Extragalactic Astronomy, Department of Physics, Durham University, South Road, Durham, DH1 3LE, UK\\
$^{32}$ Observatoire de Sauverny, Ecole Polytechnique F\'ed\'erale de Lau- sanne, CH-1290 Versoix, Switzerland\\
$^{33}$ Dipartimento di Fisica e Astronomia, Universit\'a di Bologna, Via Gobetti 93/2, I-40129 Bologna, Italy\\
$^{34}$ INFN-Bologna, Via Irnerio 46, I-40126 Bologna, Italy\\
$^{35}$ Department of Astronomy, University of Geneva, ch. d'\'Ecogia 16, CH-1290 Versoix, Switzerland\\
$^{36}$ AIM, CEA, CNRS, Universit\'{e} Paris-Saclay, Universit\'{e} Paris Diderot, Sorbonne Paris Cit\'{e}, F-91191 Gif-sur-Yvette, France\\
$^{37}$ Space Science Data Center, Italian Space Agency, via del Politecnico snc, 00133 Roma, Italy\\
$^{38}$ INFN-Sezione di Bologna, Viale Berti Pichat 6/2, I-40127 Bologna, Italy\\
$^{39}$ Instituto de Astrof\'isica e Ci\^encias do Espa\c{c}o, Faculdade de Ci\^encias, Universidade de Lisboa, Tapada da Ajuda, PT-1349-018 Lisboa, Portugal\\
$^{40}$ Departamento de F\'isica, Faculdade de Ci\^encias, Universidade de Lisboa, Edif\'icio C8, Campo Grande, PT1749-016 Lisboa, Portugal\\
$^{41}$ Universidad Polit\'ecnica de Cartagena, Departamento de Electr\'onica y Tecnolog\'ia de Computadoras, 30202 Cartagena, Spain\\
$^{42}$ Infrared Processing and Analysis Center, California Institute of Technology, Pasadena, CA 91125, USA\\
}

   \date{Received: 18 February 2020 / Accepted: 26 October 2020}

% \abstract{}{}{}{}{} 
% 5 {} token are mandatory
 
  \abstract
   {The ESA \textit{Euclid} space telescope could observe up to 150\,000 asteroids as a side product of its primary cosmological mission. Asteroids appear as trailed sources, that is streaks, in the images. Owing to the survey area of 15\,000 square degrees and the number of sources, automated methods have to be used to find them. \textit{Euclid} is equipped with a visible camera, VIS (VISual imager), and a near-infrared camera, NISP (Near-Infrared Spectrometer and Photometer), with three filters.}
   {We aim to develop a pipeline to detect fast-moving objects in \textit{Euclid} images, with both high completeness and high purity.}
   {We tested the \texttt{StreakDet} software to find asteroids from simulated \textit{Euclid} images. We optimized the parameters of \texttt{StreakDet} to maximize completeness, and developed a post-processing algorithm to improve the purity of the sample of detected sources by removing false-positive detections.}
   {\texttt{StreakDet} finds 96.9\% of the synthetic asteroid streaks with apparent magnitudes brighter than 23rd magnitude and streak lengths longer than 15 pixels ($10\,{\rm arcsec\,h^{-1}}$), but this comes at the cost of finding a high number of false positives. The number of false positives can be radically reduced with multi-streak analysis, which utilizes all four dithers obtained by \textit{Euclid}.}
   {\texttt{StreakDet} is a good tool for identifying asteroids in \textit{Euclid} images, but there is still room for improvement, in particular, for finding short (less than 13 pixels, corresponding to 8$\,{\rm arcsec\,h^{-1}}$) and/or faint streaks (fainter than the apparent magnitude of 23).}

   \keywords{Minor planets, asteroids: general --
                Methods: numerical --
                Methods: data analysis --
                Techniques: image processing --
                Surveys
               }

    \authorrunning{M. Pöntinen et al.}
    %\titlerunning{}
   \maketitle
%
%________________________________________________________________

\section{Introduction}

European Space Agency's (ESA) upcoming, cosmological \textit{Euclid} mission surveys a large portion of the sky \citep{laureijs2011}. Even though it purposefully points further than 15\textdegree{} from the ecliptic plane, \textit{Euclid} also detects a large number of Solar System objects (SSO), up to 150\,000 \citep{carry2018}. As the telescope of \textit{Euclid} points at a fixed position of the sky during a measurement, the SSOs move relative to the background sky, and objects moving faster than $\sim5\,{\rm arcsec\,h^{-1}}$ (from near-Earth asteroids to Jupiter Trojans) appear as streaks in the data. These fast-moving objects form the majority, approximately two-thirds, of SSOs detected by \textit{Euclid} (see Table \ref{table:SSO}). Thus, the main problem to be solved is streak detection. The challenges are that the images contain many other sources, such as stars and galaxies, and that most of the linear features in the images are due to cosmic rays rather than asteroids.

The importance of finding the asteroids in \textit{Euclid} images is twofold. First, they offer valuable data for Solar System science. Most of the goals for studying asteroids benefit from understanding their compositions \citep{gaffey2002}. Observing the spectral energy distribution of asteroids, in particular in the near-infrared, is essential for their compositional modeling \citep{reddy2015}. The \textit{Euclid} mission substantially increases the number of asteroids with multi-band photometry that extends to near-infrared. Second, the asteroids need to be identified so that they do not appear as artifacts and interfere with the cosmological data-analysis pipeline, which aims at a very precise measurement of galaxy shapes to determine the amount of weak lensing.

The first step in the analysis of \textit{Euclid} data, in terms of Solar System science, is to find the asteroids in \textit{Euclid} images, because most of the asteroids that are visible in the images are previously undiscovered objects (see Table \ref{table:SSO}). \citet{lieu2019} carried out tests using deep convolutional neural networks (CNNs) to find the streaks in \textit{Euclid} images. A related method has been developed by \citet{duev2019} to use an ensemble of CNNs to remove false-positive streaks in Zwicky Transient Facility data. The CNN approach is auspicious; however, currently it has only been used to classify small (up to a few hundred pixels wide) sub-images containing streak candidates, but not to extract streak coordinates from large images. Furthermore, even though the simulated \textit{Euclid} images imitate the future real data as closely as possible, the real images might look different. Because of the gap between synthetic and real image distributions, a CNN that is only trained with simulated training data probably does not work optimally out of the box for the real images \citep{shrivastava2017}. If machine learning is used, it is advisable to have other algorithms for finding the asteroids, if for nothing else, at least for gathering non-synthetic training examples for the neural networks. There is a parallel effort to develop a non-machine-learning pipeline from simulated \textit{Euclid} images by Nucita et al. (in prep.), which is focused on finding streaks that are shorter than 15 pixels ($10\,{\rm arcsec\,h^{-1}}$). Other astronomical streak detection methods are, for example, \texttt{findStreaks} \citep{waszczak2017} and a method based on fast Radon transform \citep{nir2018}.

\begin{table*}%[h!]
\caption{Estimates of \textit{Euclid} survey parameters for different kinds of Solar System objects. NEA stands for near-Earth asteroid, MC for Mars-crossing, MB for main-belt, Trojan for Jovian Trojan, and KBO for Kuiper-belt object. Column 4 (\textit{Euclid} observations) shows the estimated total number of Solar System objects in the data, while Column 3 (\textit{Euclid} discoveries) shows the estimated number of previously unknown observed objects. Columns 5, 6, and 7 ($H_V$ magnitude limits) show the largest absolute magnitudes $H_V$, for which the observation probability is 100\%, 50\%, and 1\%, respectively. The values were determined from simulations, and they show the likelihood that an object of a given absolute magnitude appears in \textit{Euclid} images so that the apparent magnitude is enough to observe the object. Column 8 (${\rm arcsec\,h^{-1}}$) shows the median sky motion of the objects, with 25th and 75th percentiles in the subscripts and superscripts, and Column 9 (VIS pixels) shows the corresponding streak lengths on the VIS CCD. The data are from \citet{carry2018} except for Column 2, for which the numbers of currently known objects were updated by the Minor Planet Center, as of October 17, 2020.}
\label{table:SSO}
\centering
\begin{tabular}{ccccccc}

 \hline\hline 
 \vspace{2pt}
 Population & Known objects & \textit{Euclid} discoveries & \textit{Euclid} observations & $H_V$ magnitude limits & ${\rm arcsec\,h^{-1}}$ & VIS Pixels \\
 \hline
 \vspace{4pt}
 NEA & 24\,086 & $1.4_{-0.5}^{+1.0} \times 10^4$ & $1.5_{-0.6}^{+1.0} \times 10^4$ & 22.75 \quad 23.75 \quad 26.50 & $43.3_{-19.9}^{+36.5}$ & $67.9_{-31.2}^{+57.2}$  \\
 \vspace{4pt}
 
 MC & 16\,662 & $1.0_{-0.8}^{+1.7} \times 10^4$ & $1.2_{-0.8}^{+1.7} \times 10^4$ & 21.00 \quad 21.25 \quad 22.75 & $41.3_{-14.9}^{+22.6}$ & $64.8_{-23.3}^{+35.4}$  \\
 \vspace{4pt}

 MB & 937\,295 & $8.2_{-2.2}^{+2.5} \times 10^4$ & $9.7_{-2.2}^{+2.5} \times 10^4$ & 19.50 \quad 20.00 \quad 21.25 & $32.5_{-5.5}^{+7.9}$ & $51.0_{-8.6}^{+12.4}$  \\
 \vspace{4pt}

Trojan & 8846 & $7.1_{-4.9}^{+9.3} \times 10^3$ & $7.5_{-5.0}^{+9.5} \times 10^3$ & 17.00 \quad 17.25 \quad 18.25 & $13.3_{-1.1}^{+1.4}$ & $20.9_{-1.7}^{+2.2}$  \\
\vspace{4pt}

Centaur & 951 & $2.2_{-1.4}^{+2.1} \times 10^3$ & $2.2_{-1.4}^{+2.1} \times 10^3$ & 14.75 \quad 15.50 \quad 18.25 & $4.0_{-1.5}^{+2.9}$ & $6.2_{-2.3}^{+4.5}$  \\
\vspace{4pt}

KBO & 2553 & $5.3_{-1.3}^{+1.6} \times 10^3$ & $5.5_{-1.3}^{+1.6} \times 10^3$ & \phantom{1}8.25 \quad \phantom{1}8.75 \quad 10.00 & $0.6_{-0.1}^{+0.3}$ & $1.0_{-0.2}^{+0.5}$  \\
 \vspace{4pt}

Comet & 4070 & $22_{-4}^{+4}$ & $38_{-4}^{+5}$ & 18.25 \quad 19.00 \quad 22.00 & $4.4_{-1.8}^{+6.2}$ & $6.9_{-2.8}^{+9.7}$  \\
 
\hline
\end{tabular}
\end{table*}

The method used in this paper to tackle the problem of finding the asteroids, especially the faster-moving ones, is a software called \texttt{StreakDet}, which was developed by \citet{virtanen2016} to detect streaks caused by space debris in images obtained from an Earth-orbiting platform. We tested what proportion of simulated SSOs can be detected from the synthetic images with \texttt{StreakDet}, what the detectable apparent magnitude and sky motion ranges are, and how high completeness and purity can be achieved. We did this by optimizing the parameters of \texttt{StreakDet}, developing a post-processing step to decrease the number of false positives, and programming a test and analysis software to give statistics on the results. The simulated data were generated with \texttt{ELViS}, a \texttt{Python} program developed within the Euclid Consortium.

In what follows, we first describe the \textit{Euclid} mission and its predicted main contributions to Solar System science. Then, we describe the simulation of \textit{Euclid} VIS images, the methods for extracting streaks from images, and the tools for analyzing the results. We then present and discuss our results obtained with the \texttt{StreakDet} software and our additional post-processing step. Finally, we provide our conclusions.

\section{\textit{Euclid}}

\subsection{\textit{Euclid} mission}

\textit{Euclid} is a medium-class mission in the Cosmic Vision program of ESA. It was selected in October 2011 and is planned to launch in 2022. The nominal science program duration is six years, a time during which \textit{Euclid} stays in a halo orbit around the second Lagrange point of the Sun-Earth system. The mission is named after the mathematician of ancient Greece, Euclid of Alexandria.

\textit{Euclid} is a cosmology mission, and it surveys a large portion of the sky, 15\,000 deg\textsuperscript{2} \citep{laureijs2011}. \textit{Euclid} explores the nature of dark energy, dark matter, and gravity using gravitational lensing effects on galaxies (weak lensing) and the properties of galaxy clustering (baryonic acoustic oscillations and redshift space distortion) \citep{amendola2018}.

\textit{Euclid} has a telescope with an aperture of 1.2 meters and a focal length of 24.5 meters \citep{venancio2014}. The mission operates in the wavelength range from 550\,nm (green) to 2000\,nm (near-infrared). The measuring instruments are VIS (VISual imager) and NISP (Near-Infrared Spectrometer and Photometer), sharing the same 0.53 deg\textsuperscript{2} field of view. VIS is a 600-megapixel visible imager, operating between the wavelengths of 550 and 900\,nm \citep{cropper2018}. NISP consists of a near-infrared three-filter photometer (NISP-P) and a slitless spectrograph (NISP-S) with a resolving power of 380 \citep{maciaszek2016}. The pixel size for VIS is 0.1 arcseconds per pixel, and for NISP 0.3 arcseconds per pixel. \textit{Euclid} operates in a step-and-stare mode: VIS and NISP-S observe an area of the sky simultaneously with an exposure time of 565 seconds, after which NISP-P executes three shorter exposures with $Y$, $J$, and $H$ filters for 121\,s, 116\,s, and 81\,s, respectively \citep{racca2016}. For a given field, the aforementioned exposures are repeated four times, with small changes in the telescope pointing direction in between them, so that the total observing time for each field is approximately 70 minutes. The current baseline now also includes a fifth VIS image, taken simultaneously with the NISP/$J$ photometry exposure, during the third jitter.

Since the data created by \textit{Euclid} consist of hundreds of thousands of images, with an accumulated data volume of several petabytes, automated algorithms are necessary to analyze the data. Approximately 10 billion galaxies appear in the data, of which over 1 billion are used for weak-lensing, and several tens of millions of galaxy redshifts are measured \citep{laureijs2011}.

\subsection{Solar System objects in \textit{Euclid} images}

Although \textit{Euclid}'s main science goals are cosmological, it also observes up to 150\,000 Solar System objects \citep{carry2018}. Most of these are asteroids, and the ones faster than $\sim5\,{\rm arcsec\,h^{-1}}$ (orbits approximately up to Jupiter's orbit) appear in the \textit{Euclid} images as streaks. The detection limit with VIS for $10\sigma$ on 0.43\,arcsec extended sources is around $m\textsubscript{AB} = 24.9$ \citep{cropper2018}, and with NISP $Y$, $J$, and $H$ filters for $5\sigma$ point sources, it is around $m\textsubscript{AB} = 24$. The NISP slitless spectra is obtained by two grisms, blue (0.92 to 1.25\,µm), which is used only in the deep survey, and red (1.25 to 1.85\,µm), which is used both in wide and deep surveys. The grisms have a continuum sensitivity of around $m\textsubscript{AB} = 21$. 

Detailed estimates of the observed SSOs are shown in Table \ref{table:SSO}. In addition to the object classes shown in Table \ref{table:SSO}, \textit{Euclid} has the potential to observe interstellar objects (ISO) as well. The Vera C.\ Rubin Observatory (VRO; formerly known as the Large Synoptic Survey Telescope) is estimated to find on the order of ten, possibly even more, ISOs during its ten-year survey \citep{cook2016}. \textit{Euclid} covers a slightly smaller area of the sky than VRO, and the survey is shorter, so as a crude estimate, the number of detected ISOs is expected to be a bit smaller than VRO. All in all, \textit{Euclid} can make a viable contribution to the detection of ISOs, especially in the northern sky, because VRO covers the southern sky only.

\textit{Euclid} avoids the Galactic and ecliptic planes by observing the sky that has Galactic latitudes greater than 30\textdegree, and ecliptic latitudes higher than 15\textdegree, except for calibration fields, which can be closer to the ecliptic plane. For this reason, the asteroids detected by \textit{Euclid} are primarily objects in high-inclination orbits, which are currently underrepresented in the census of asteroids \citep{mahlke2018}. Nevertheless, the calibration field data close to the ecliptic plane contain the highest number of asteroids per field of view, up to a few thousand. There are approximately 60\% more known asteroids for every 3 degrees closer to the ecliptic. There are also some asteroids in the deep fields, but not many because they are far away from the ecliptic.

As \textit{Euclid} measures multi-band photometry of galaxies, it also substantially increases the number of asteroids with multi-band photometry in the near-infrared region. In addition to the compositional information provided by multi-band photometry, \textit{Euclid} data can, in many cases, also be used to constrain several other properties of the asteroids, such as the rotation period, spin-axis orientation, and shape, as well as detect binary asteroids \citep{carry2018}. Due to the relatively short observational time span per asteroid, accurate orbit solutions for the discoveries cannot be obtained by using \textit{Euclid} data only, but rough orbits and especially inclination distributions can be estimated \cite[cf.][]{muinonen2006}. More accurate orbits can be obtained if additional astrometry is or becomes available. For example, the upcoming VRO determines the orbits for a large number of asteroids, and these orbits can be cross-correlated with \textit{Euclid} detections \citep[cf.][]{Snodgrass2018}.

\section{Simulated \textit{Euclid} images and preprocessing}

The simulated data are generated with \texttt{ELViS}, a \texttt{Python} program developed within the Euclid Consortium, to mimic the actual \textit{Euclid} VIS data as closely as possible. Here, we focus on the simulated VIS data, as it is the instrument with the higher sensitivity, longer exposure time, and smaller pixels, thus being better suited for the \texttt{StreakDet} software. Simulated VIS images include astronomical sources such as galaxies, stars, and Solar System objects as well as image artifacts such as cosmic rays (CRs), ghosts, charge transfer inefficiency (CTI), and bleeding. The images also contain Poisson noise, readout noise, and bias. See the top left image of Fig.~\ref{fig:Steps} for an example of the data.

The field we simulated is centered at Galactic latitude 85.0\textdegree{} and longitude $-82.8$\textdegree{}.The number of galaxies per CCD range from 469 to 822, with a mean of 653. For stars, the range is from 78 to 126, and the mean is 100 stars per CCD. The simulation galaxy input catalog was created by the \textit{Euclid} cosmology science working group based on the MICE2 simulation \citep{crocce2015}, for the usage of \textit{Euclid} "Science Challenge 3". Due to undocumented reasons, probably related to processing volume, the galaxy count is approximately half of what is expected for the real sky, and galaxies fainter than 24.5 magnitudes are left out. 

The asteroids in our data set were simulated to come from a uniform random distribution between apparent magnitudes 20 and 26, apparent velocities between 10 and $80\,{\rm arcsec\,h^{-1}}$, and angles between 0 and 360 degrees (clockwise from east), except for one image set that focused more on shorter streaks, with velocities ranging from 1 to $20\,{\rm arcsec\,h^{-1}}$. On average, there are 25 simulated asteroid streaks in each CCD. The simulated SSO population is in no way realistic, but this flat synthetic population is more suited for analyzing the performance metrics of \texttt{StreakDet} and our pipeline.

Due to a reason that has yet to be identified within the \texttt{ELViS} software, there is a small systematic, visible offset (up to a few pixels) between the intended coordinates of the simulated asteroids and the final positions of the simulated asteroid streaks. The offset is a linear function of right ascension and declination. Therefore, we applied a linear correction to the ground-truth catalogs to correct for the systematic offset.

The simulated data are provided as multi-extension FITS files, and each file contains a header and four 2048\,$\times$\,2066-pixel quadrants of a single 4096\,$\times$\,4132-pixel CCD. One simulated exposure consists of 36 such CCDs, forming a 6\,$\times$\,6 grid. Between exposures, there is a dither movement, and one data set consists of four dithered exposures, totaling 144 CCDs. The simulated data did not include the fifth shorter VIS exposure. We run tests on ten such sets for faster moving objects, and one set for slower objects, totaling 11 sets, adding up to a total of 1584 CCDs.

Before feeding the images to \texttt{StreakDet}, we tiled the 2\,k\,$\times$\,2\,k-pixel image quadrants of multi-extension FITS files into single-extension 4\,k\,$\times$\,4\,k-pixel images. We also removed CRs with \texttt{Astro-SCRAPPY} \citep{mccully2018, vandokkum2001}. \texttt{Astro-SCRAPPY} removed essentially all bright pixels caused by cosmic rays. Some residuals were visible in the images afterward, which was mainly caused by the CTI effect (see top right image in Fig.~\ref{fig:Steps}). \texttt{Astro-SCRAPPY} does not remove asteroid streaks due to their different PSF shape compared to cosmic-ray streaks.

The data reduction for actual \textit{Euclid} images is handled by \textit{Euclid} OU-VIS (Organizational Unit VIS) with more advanced pipelines than were used in this work. The OU-VIS reduction pipeline includes bias subtraction, flat fielding, CR removal, and CTI corrections, as well as astrometric and photometric calibrations with the Gaia \citep{gaia2016} catalog \citep{dubath2016}.

Most of the asteroids appear in three or four dithers of the simulated data. Exceptions include asteroids that are close to the edge of an image in first dither, which subsequently move out of the imaging field due to their apparent velocity or the dither movement of the imaging field. Similarly, asteroids that are outside the imaging area in the earlier dithers can move into the images in later dithers. Thus, some asteroids appear in as little as one dither. For our simulated data, in which the apparent velocities, magnitudes, and streak angles come from a uniform random distribution, the visibility of asteroids was as follows: 37\% for four streaks, 45\% for three streaks, 9\% for two streaks, and 9\% for one streak. For a realistic asteroid population distribution, the visibility statistics are likely to differ. Nevertheless, for most objects, it is not necessary to find all the streaks in all dithers caused by an object, but as little as one out of four can count as a find, or as little as two when utilizing the post-processing step to link streaks from several dithers.

\section{Methods}

\subsection{StreakDet}

\begin{figure*}%[h]
    \centering
    \includegraphics[width=\textwidth]{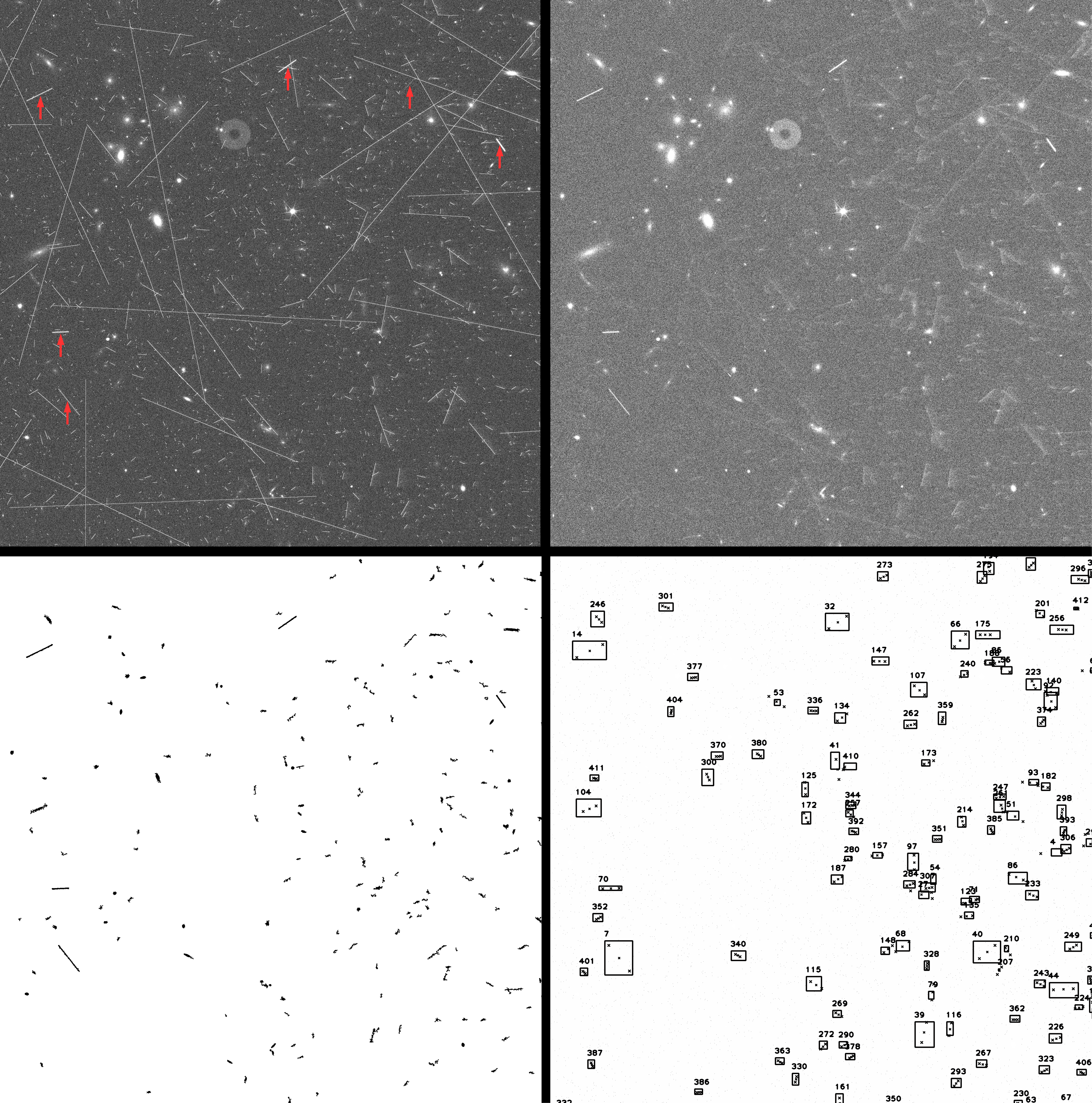}
    \caption{Streak detection steps. The \emph{top left} image shows a quadrant of a raw CCD file. The size of the image is 2048\,$\times$\,2066 pixels, corresponding to 204.8\,$\times$\,206.6 arcseconds. Asteroids are marked with red arrows, and the remaining streaks are cosmic rays. The \emph{top right} image is the same file after having cosmic rays removed. This image is then fed to \texttt{StreakDet}, which results in a segmented image, shown in the \emph{bottom left}. The \emph{bottom right} image shows the final streaks found by \texttt{StreakDet}. As can be seen, there are many false-positive detections. The brightest pixels in an image are often caused by cosmic rays, and after they are removed, the scaling of the image changes. For this reason, after cosmic ray removal, the image appears brighter than before the cosmic rays were removed.}
    \label{fig:Steps}
\end{figure*}

\texttt{StreakDet} \citep{virtanen2016} is an ESA-funded software developed to detect and analyze object trails in imaging data. \texttt{StreakDet} has been developed mainly to analyze observations of space debris either from Earth-based or space-based platforms, and it can detect long, faint, and also curved streaks. Its focus is on being able to detect streaks from single images as opposed to finding consecutive streaks from stacked images, a task for which it is not well optimized. The initial tests run (on non-\textit{Euclid} images) by the \texttt{StreakDet} developers gave detection sensitivities of about 90\% for bright streaks (per pixel signal-to-noise ratios of >1), and about 50\% for dimmer streaks ($\rm S/N = 0.5$).

The \texttt{StreakDet} pipeline consists of three main phases: segmentation, classification, and lastly astrometric and photometric reduction. The following descriptions of the algorithms are summarized from \citet{virtanen2016}. In the following description, black and white (BW) refers to a binary image with a color depth of 1 bit, where every pixel has either a value of 0 or 1. Grayscale (GS) refers to the original image with a higher bit depth.

The segmentation step converts the analyzed image into BW to make the following steps computationally less demanding. The BW image is created with the aid of two GS mean filters. The first filter uses a smaller area, for example, 3\,$\times$\,3 pixels, for which it calculates a mean pixel value. The second filter uses a larger area, for example, 21\,$\times$\,21 pixels, and calculates the mean pixel value for that area. The BW image is created by subtracting the differences of the means calculated by the GS mean filters. The idea behind the mean-differences is to detect groups of pixels whose value differs from the background. Because the mean-difference calculation is locally executed, it is typically not biased by global background gradients, which reduces the need for preprocessing the image before feeding it to \texttt{StreakDet}. Before segmentation, the density of bright pixels in the image is calculated or manually set in order to determine the proportion of white pixels to black pixels in the segmented image.

After the segmentation has turned the image to black and white, filtering processes are applied, which aim to remove non-streak-like features from the pipeline. An adapted version of binary erosion is used, which gets rid of isolated active pixels and keeps pixels which are part of a larger structure, such as a streak. Then a reconstruction filter is used to strengthen the remaining features. After the previous steps have removed most of the noise and small stars, larger stars are removed by multiple-window pixel-density evaluation, which removes pixel groups whose number of active pixels does not grow linearly when the window size is increased. Finally, the remaining features and their properties are indexed into a list with a connected component labeling (CCL) algorithm.

After segmentation, the CCL features are given as input into the classification phase.\ This consists of the following three steps: first, the characterization of BW CCL features; second, the characterization of BW CCL features that correspond to streaks; third, the characterization of original GS features that correspond to streaks.

Each of the steps uses classification processes to find the streaks, and filtering processes to eliminate the non-streaks. During each step, eigenvalue analysis is used to compute feature parameters such as width, orientation angle, aspect ratio, curvature, and porosity (referring to the compactness of the feature). Both linking and unlinking are done during the BW steps. Linking implies connecting found features that are likely to be parts of the same longer streak. In contrast, unlinking implies dividing a large found feature into smaller ones, if it seems likely that the sub-features are actually separate streaks. During the GS parametrization, point-spread-function (PSF) fitting with a moving 2D Gaussian approach, utilizing the nonlinear Levenberg-Marquardt least-squares method, is used to refine the streak parameters. The GS parametrization is done by starting with the BW features and then finding the parameters of the corresponding GS streak with the aforementioned algorithms. The GS fitting also extends outside the BW bounding box, in case only a part of the real streak was found during the BW process. Finally, the GS features are classified and filtered according to their PSF width and curvature. An example of the results after segmentation and after the whole pipeline can be seen in the bottom left and bottom right images of Fig.~\ref{fig:Steps}, respectively. The final, optional phase of the \texttt{StreakDet} pipeline is astrometric and photometric reduction.

Because \texttt{StreakDet} uses only one CPU most of the time, $N$ \texttt{StreakDet} processes can be run in parallel with a computer with $N$ CPU cores, which makes it possible to run \texttt{StreakDet} on 36 FITS CCDs in only approximately $\max(1,\;36\,N^{-1})$ times the time it takes to analyze one FITS CCD. With our settings, running \texttt{StreakDet} on a 4\,k\,$\times$\,4\,k image using a single CPU core took approximately one minute. It takes \textit{Euclid} 70 minutes to produce the 180 VIS CCDs of one pointing, so, theoretically, three CPU cores would be enough to run \texttt{StreakDet} with VIS data so that the program would not fall behind the image delivery pace of \textit{Euclid}. In practice, more CPUs are needed to rerun \texttt{StreakDet} with different parameters or on differently preprocessed VIS images or, alternatively, depending on the schedule at which VIS data become available.

\subsection{Analysis program} \label{analysis_program}

In order to compare the results given by \texttt{StreakDet} to the simulation inputs (ground truth), we developed a test and analysis program in \texttt{Python}. The analysis program was used to process all the ground-truth and \texttt{StreakDet} data, compute statistics on the results, and show plots of hits, misses, false positives, and errors.

We tested \texttt{StreakDet} by running it on single 4\,k\,$\times$\,4\,k-pixel FITS CCDs, and only afterward did we combine and stack all the data by the separate \texttt{Python} analysis program. The reason for this is that \texttt{StreakDet} itself does not appear to work well with stacked images, that is, the detection percentage was lower for stacked images than for individual images. The reason for which seems to be due to combined noise, causing the signal-to-noise ratio (S/N) of the streaks to decrease. The 4\,k\,$\times$\,4\,k images are run one by one with \texttt{StreakDet}, and then the images, \texttt{StreakDet} results, and ground-truth data are tiled and stacked afterward with the analysis program.

\begin{figure}%[h]
    \centering
    \includegraphics[width=\hsize]{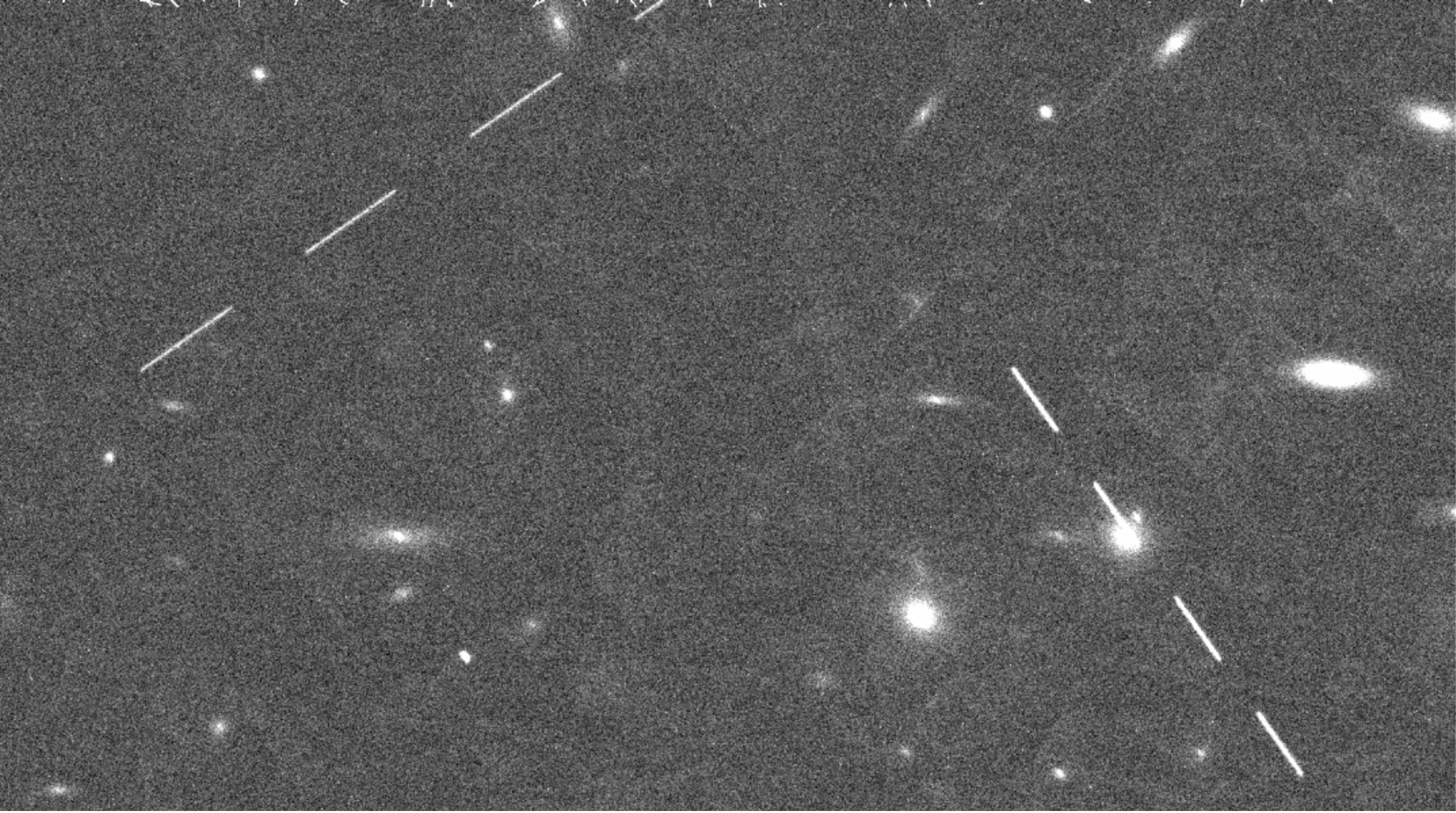}
    \caption{Example of asteroids appearing as multi-streaks, when all four dithers are stacked into a single image. The image shows two dashed lines, each of which is caused by an asteroid. The width and height of the image are approximately 1000 and 600 pixels, corresponding to 100 and 60 arcseconds, respectively.}
    \label{fig:multi-streak}
\end{figure}

To improve the purity of the sample (i.e., to decrease the number of false-positive detections), we developed a post-processing step called multi-streak analysis and implemented it in \texttt{Python}. As the data have images from four dithers, the asteroids appear as multiple line segments along the same line in the stacked image (see Fig.~\ref{fig:multi-streak}). Before feeding the data to the multi-streak pipeline, so-called static streaks are removed. In other words, streaks that are in the same coordinates in multiple dithers cannot be asteroids, but they are galaxies or other static objects instead and thus can be removed from the analysis. With our optimized \texttt{StreakDet} parameters, approximately one quarter of all streaks found by \texttt{StreakDet} were static.

The multi-streak pipeline is described below.
\begin{enumerate}
\item Run \texttt{StreakDet} separately on all FITS CCDs for all four dithers.
\item Tile and stack ground-truth data with the \texttt{Python} analysis program.
\item Tile and stack \texttt{StreakDet} results with the \texttt{Python} analysis program.
\item Search for ground-truth multi-streaks that have 2--4 streaks along the same line, so that the streaks are different dithers in logical order, that is, assuming that the object moves fairly linearly in one direction and does not suddenly move backward.
\item Search for \texttt{StreakDet} multi-streaks that have 2--4 streaks along the same line (again, all streak in different dithers should be in logical order).
\item Check that the lengths, angles, fluxes, and PSF sigma (standard deviation from PSF fitting) values of the individual streaks in the multi-streaks are within given ranges of each other.
\item Analyze which \texttt{StreakDet} multi-streaks match ground-truth multi-streaks and which do not. To be classified as a match, at least two single streaks in a \texttt{StreakDet} multi-streak have to match to those of the ground-truth multi-streak.
\end{enumerate}
Without the ground-truth data available, that is, when processing the real \textit{Euclid} data, only steps 1, 3, 5, and 6 are carried out.

Asteroids that move from one CCD to another between dithers are taken into account in the multi-streak analysis. For example, if an asteroid appears in one CCD in the first dither, and moves to the neighboring CCD in the subsequent dithers, the multi-streak analysis can link all of the separate streaks together.

\section{Results}

\subsection{Single-streak analysis}

% Two column figure (place early!)
\begin{figure*}%[h]
    \centering
    \includegraphics[width=\textwidth]{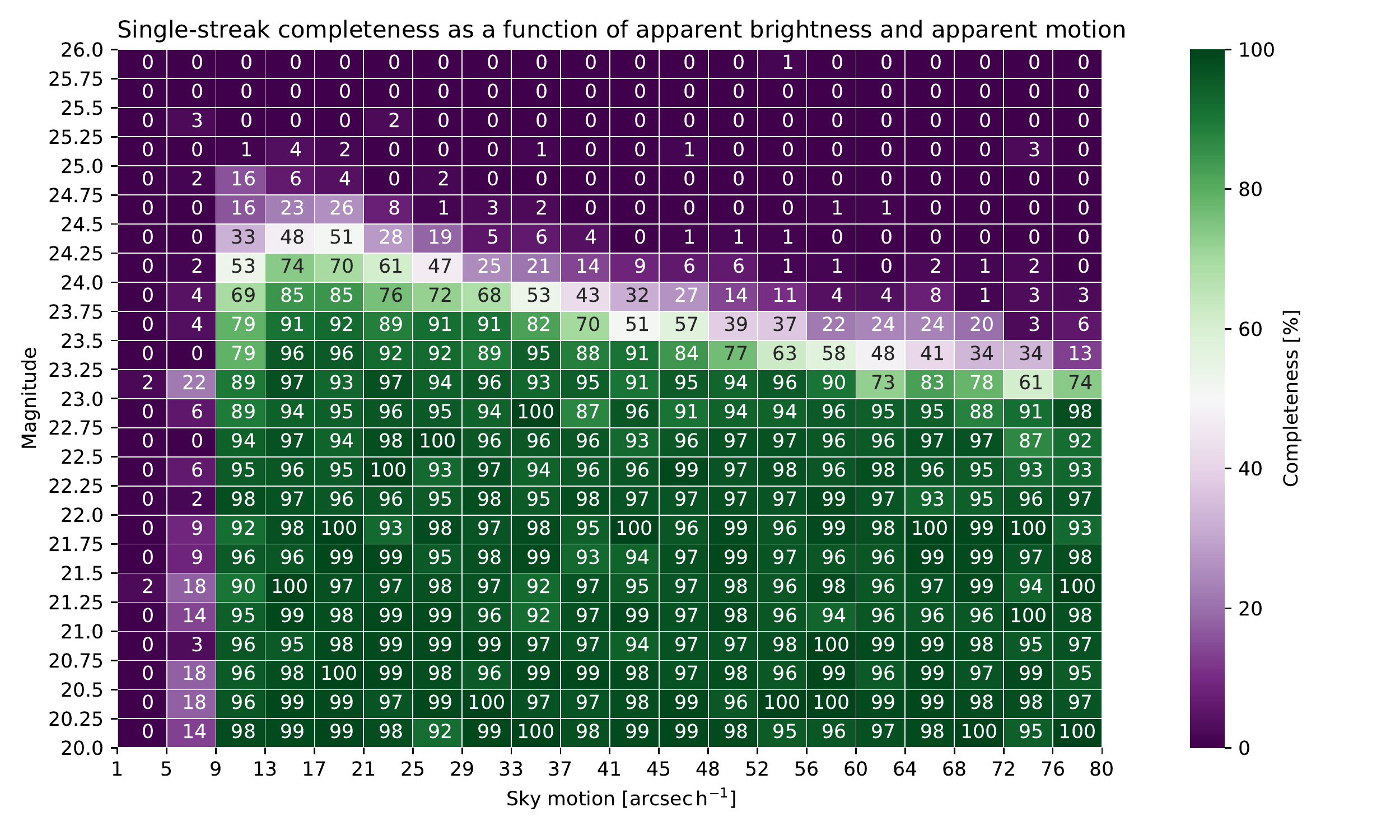}
    \caption{\texttt{StreakDet} detection percentage as functions of apparent magnitude and apparent motion. The values on both axes show the limits of the bins. For example, the bin between sky motions of 33 and 37 $\,{\rm arcsec\,h^{-1}}$ and magnitudes of 21.5 and 21.75 shows that the \texttt{StreakDet} detection percentage (completeness) is 99\% for streaks created by all simulated objects between those values. On average, there are 83 ground-truth streaks per bin.}
    \label{fig:Heatmap}
\end{figure*}

For \texttt{StreakDet}, the fraction of true positives and especially false positives changes substantially depending on the settings used. After testing the parameters available for the \texttt{StreakDet} pipeline, we identified a few parameters that increase the number of true positives. In general, the segmentation settings are the most important part of optimizing \texttt{StreakDet}, because if a streak is not already found in the segmentation phase, it becomes impossible to find it during the subsequent processing stages. After optimizing the segmentation phase, parameters related to the later stages of the pipeline were optimized as well in order to get the final detection percentage as close to the segmentation detection percentage as possible. After testing and optimization, a combination of settings appearing to maximize the number of true positives was chosen and used to process all simulated \textit{Euclid} images.

As expected, there is a trade-off between maximizing the number of true positives and minimizing the number of false positives. Our primary goal was to maximize the number of true positives during \texttt{StreakDet} processing because the number of false positives can be radically reduced with the multi-streak analysis that is carried out as a post-processing step. The optimal parameters seem to vary slightly from image to image, and they may depend on the exact image characteristics, such as background brightness and stellar density. However, \texttt{StreakDet} takes the number of bright stars and pixels in the image into account before starting the segmentation, so the variation in results between images is small, at least with the simulated data. It is expected that \texttt{StreakDet} can achieve fairly similar results on different parts of the sky, using the same parameters. When aiming for the highest completeness on atypical fields, such as the calibration fields, tuning the parameters might prove useful.

\texttt{StreakDet} can differentiate between streaks caused by asteroids and CRs due to their difference in PSF shape. However, when tested on data containing CRs, the total detection percentage achieved by \texttt{StreakDet} was approximately 10 percentage points lower compared to data that had CRs removed. Due to this effect, we opted to remove CRs from the data before analyzing it with \texttt{StreakDet}. Furthermore, the CRs are removed anyway during the processing of the real \textit{Euclid} data by OU-VIS.

\begin{figure*}%[h]
    \centering
    \includegraphics[width=\textwidth]{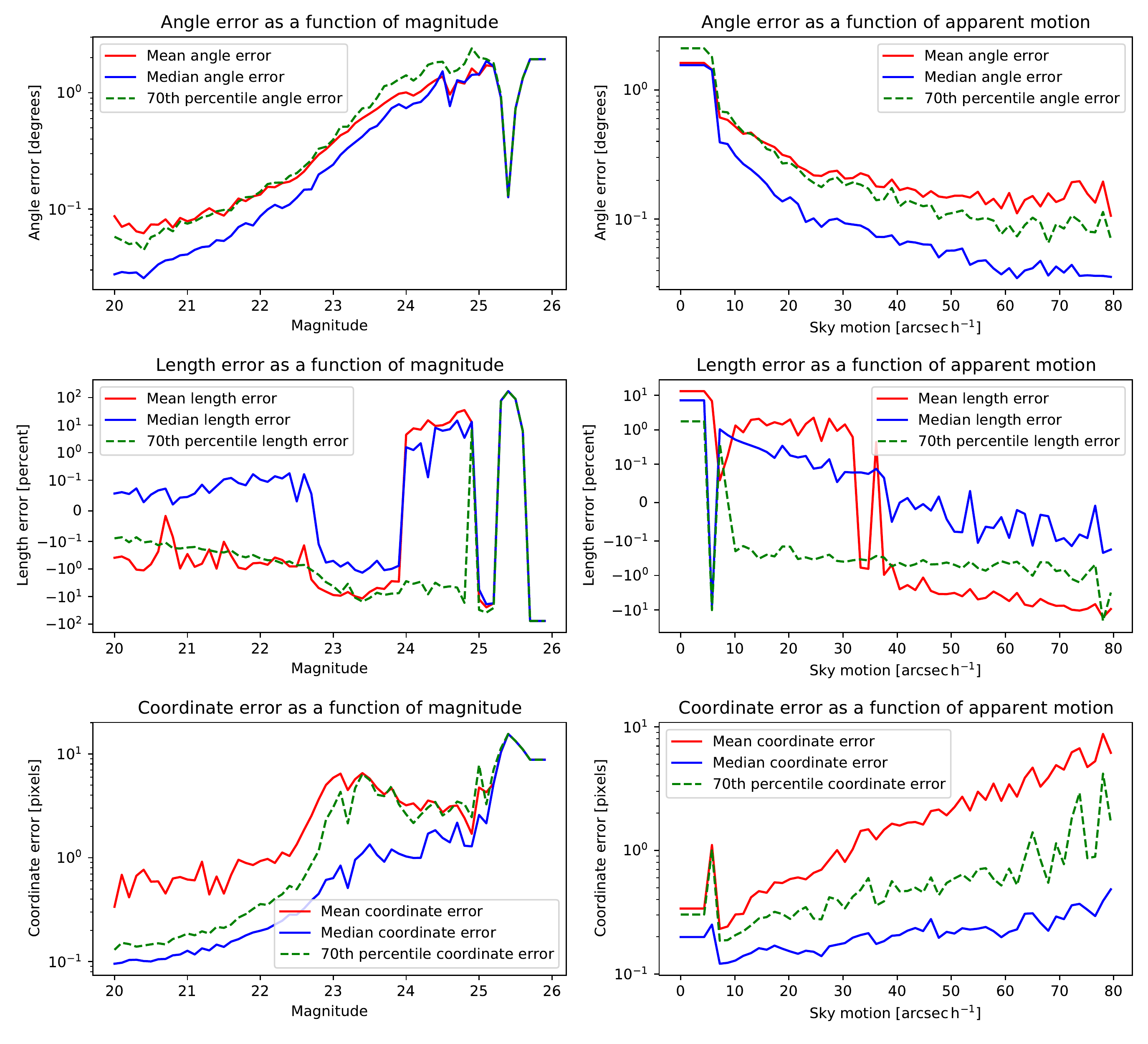}
    \caption{Errors in angles, lengths, and coordinates of \texttt{StreakDet} finds as a function of magnitude and length. The y-axes of the two uppermost and the two lowermost plots are in logarithmic scale. The y-axes of the middle plots are in symmetric logarithmic scale, where the values between $-10^{-1}$ and $10^{-1}$ are shown in linear scale. In the middle plots, showing length errors, negative y-values mean that the corresponding lengths given by \texttt{StreakDet} are shorter than ground-truth lengths, whereas positive values mean that lengths given by \texttt{StreakDet} are longer than ground truth. The 70th percentile length errors in the middle plots are technically 30th percentile errors, since the length errors are typically negative.}
    \label{fig:Errors}
\end{figure*}

Figure~\ref{fig:Heatmap} shows the \texttt{StreakDet} detection percentage (completeness) for streaks of different magnitudes and lengths as a heat map. The completeness is defined here as the fraction of streaks found among the streaks that are at least partially visible in the CCDs. In other words, streaks that are completely outside or between the CCDs are ignored. All in all, \texttt{StreakDet} managed to find 96.9\% of individual streaks brighter than 23 magnitudes and longer than 15 pixels (16\,283 out of 16\,807). For fainter streaks, the detection percentage starts to fall, reaching essentially zero at magnitude 25. This is to be expected, as the limiting magnitude in the simulated data is 24.5 at an S/N of 10 for an extended source. This approximately corresponds to a magnitude of 26.5, using an S/N of 3 to 5 for a point source. Since the light is spread onto, for example, 10 pixels, a magnitude 24 SSO has approximately the same per-pixel counts as a magnitude 26.5 star, and it becomes difficult to discern from the background. For short streaks, the detection percentage decreases rapidly for streaks shorter than 13 pixels (8$\,{\rm arcsec\,h^{-1}}$). For fast-moving faint asteroids, the streaks appear fainter than for slow-moving asteroids of the same magnitude because, in a longer streak, the integrated flux is spread over a larger number of pixels. For this reason, shorter streaks caused by faint asteroids are found more easily than longer streaks caused by faint asteroids. Since the simulated galaxy density is only half of what is excepted of the real sky and the numerous faint galaxies are missing from the images, the detection percentage for faint streaks could be somewhat different for the actual data. Nevertheless, as \texttt{StreakDet} does not detect streaks fainter than magnitude 24.5, the difference in results should not be very significant.

Each SSO forms from one to four streaks, because there are four dithers. Therefore, it is not necessary to detect each streak that formed by an SSO in order to be able to detect that particular object. When calculating the number of simulated SSOs for which \texttt{StreakDet} found at least one streak out of up to four streaks, the detection percentage is 99.1\% for SSOs brighter than 23 magnitudes and longer than 15 pixels (10$\,{\rm arcsec\,h^{-1}}$). Again, the detection percentage starts to fall after magnitude 23, but it is 10 to 20 percentage points higher for SSOs than for individual streaks between magnitudes 23 and 24.5.

According to our tests, \texttt{StreakDet}'s detection ability declines rapidly for streaks shorter than approximately 13 pixels (8$\,{\rm arcsec\,h^{-1}}$). Therefore, for the ten main image sets used for the final results, the lengths of asteroid streaks were chosen to range from 15 to 125 pixels (from 10 to 80 $\,{\rm arcsec\,h^{-1}}$). Also, with certain settings, \texttt{StreakDet} does not appear to find streaks shorter than 15 pixels (10$\,{\rm arcsec\,h^{-1}}$) and brighter than 20.5 magnitudes at all. This is due to \texttt{StreakDet} being optimized for relatively  faint streaks by default. When the pixel brightness crosses a certain threshold, it causes bright streaks to already be filtered out in the segmentation phase. This can be avoided by adjusting parameters in the source code of \texttt{StreakDet}'s segmentation phase. Optimizing for short and bright streaks causes the completeness to fall slightly, approximately 0.5 percentage points for all streaks fainter than 21 magnitudes. This is notable because most of the found asteroids are faint, and a small change in the detection percentage close to the limiting magnitude radically changes the number of observed objects. It seems that one set of parameters is good for finding the brightest streaks, and another is better for finding fainter ones. For the real \textit{Euclid} images, it might make sense to run \texttt{StreakDet} twice for all data, with two different parameter sets. For the results in this work, we used the setting optimized for brighter streaks, because it offered a slightly higher total completeness with our uniform data set.

In addition to the ten main image sets focused on streaks longer than 15 pixels (10$\,{\rm arcsec\,h^{-1}}$), we ran supplementary tests with a simulated data set containing shorter streaks, from approximately 1 pixel to 30 pixels (1--20$\,{\rm arcsec\,h^{-1}}$). We did this to get a better idea of the streak length range at which \texttt{StreakDet} can operate. According to those tests, \texttt{StreakDet}'s detection percentage starts to drop sharply for streaks shorter than 13 pixels (9$\,{\rm arcsec\,h^{-1}}$), reaching essentially zero below 9 pixels (6$\,{\rm arcsec\,h^{-1}}$).

Comparing the results in Fig.~\ref{fig:Heatmap} to the populations shown in Table~\ref{table:SSO}, in terms of sky motion, \texttt{StreakDet} can find a vast majority of bright enough near-Earth asteroids, Mars-crossing asteroids, main-belt asteroids, and Jovian Trojans \citep[for a visualization of the sky motion ranges of different populations, see Fig. 5 in][]{carry2018}. In addition, individual Centaurs, Kuiper-belt objects, and comets can form long enough streaks for \texttt{StreakDet} to detect. For the populations shown in Table~\ref{table:SSO}, the apparent magnitude limit is 24.5, which \texttt{StreakDet} can reach for the shortest found streaks (Trojans) with a completeness of approximately 50\%. For faster moving objects, such as main-belt asteroids, the limiting magnitude of \texttt{StreakDet} is 24 or less. This suggests that \texttt{StreakDet} cannot detect all of the up to 150\,000 objects appearing in the images, but getting precise completeness and purity values for realistically simulated SSO populations requires new simulations and remains for future work.

The brightness of each simulated asteroid stayed constant in all the streaks it formed. In reality, asteroids rotate and often have nonspherical shapes, which causes the apparent brightness to vary. Most of the asteroids larger than approximately 150 meters have rotation periods longer than two hours, but smaller asteroids can rotate much faster \citep{pravec2002}. Therefore, as the observation time for a pointing is 70 minutes, larger asteroids can show variation in brightness mostly between different dithers. Smaller asteroids can show significant variation even within a single dither, resulting in a streak with periodic brightness variation, and if the brightness is close to the limiting magnitude, a dashed line. \texttt{StreakDet} can link a dashed line into a coherent streak feature, but the variation in apparent magnitude can make the detection more difficult. Measuring the significance of this effect on the detection results requires more realistic simulations and remains for future work.

\subsection{Errors}

Figure~\ref{fig:Errors} shows the angle, length, and coordinate errors of \texttt{StreakDet} detections as functions of the true magnitude and true streak length. The fitted angles of the streaks are typically accurate when compared to the ground truth. For the brightest streaks, in the magnitude range of 20--21, the average angle error is around 0.06 degrees, and the median angle error is 0.03 degrees. For fainter streaks, the angle errors increase steadily due to a less accurate streak fitting, and after magnitude 24, the average error is above 1 degree. For all magnitudes, the average angle error is 0.22 degrees, and the median angle error is 0.08 degrees. After magnitude 25, there were just a few individual detections, which causes the median and mean errors to be identical. 

For shorter streaks, the angle error is larger than for long streaks due to a better streak fitting for long streaks. In the error plots that are plotted as a function of sky motion, there appears to be a plateau at the very shortest length range, but this is again due to the fact that there are very few or zero found streaks at those points.

The fitted lengths of the \texttt{StreakDet} finds are also quite accurate for bright streaks. For streaks brighter than 21 magnitudes, the average error in streak length is $-0.6$\%, that is, the streaks found by \texttt{StreakDet} are slightly (at sub-pixel level) shorter than the ground truth. The median error for magnitudes smaller than 21 is 0.04\%, so that the lengths of the streaks found are very slightly longer than the ground-truth streaks. Again, the error and its variance start to increase for fainter streaks due to less accurate streak fitting. The sign of the length error as a function of magnitude changes between magnitudes 22.6 and 24, and again around magnitude 25. When going to fainter streaks, the variance of the errors increase due to a less accurate streak fitting, so the change of the error sign is likely just an effect of randomness. This is especially true after magnitudes 24 and even more so after 25, where the number of found streaks is very small and the streaks are very faint. The very small number of detections for the faintest streaks causes the median and mean errors to become identical in the plot.

For long streaks, \texttt{StreakDet} sometimes finds two shorter line segments of the ground-truth streak instead of the entire streak. In addition, some streaks fall partially outside of the image, and only a part of them are visible on the CCD. This causes the found length to be shorter than the ground-truth length. Longer streaks are more likely to continue outside the image. These effects combined are responsible for the median and mean length errors turning negative for streaks longer than approximately 60 pixels (40$\,{\rm arcsec\,h^{-1}}$). For long streaks, the median length error is negative at sub-pixel level, whereas the mean length error approaches $-10$\%. For short streaks, the median and mean length errors are positive at sub-pixel level.

In other words, both angle and length errors increase for faint streaks. For short streaks, angle errors are larger, but relative length errors are smaller. For the long streaks, the opposite is true, so that angle errors are small, but relative length errors are larger. A notable detail is that while the mean error can be large, the corresponding median error is typically much smaller. This means that most streak parameters defined by \texttt{StreakDet} are quite close to the ground truth, and for a small number of streaks, the parameters are off by a large margin. In addition to the causes for larger errors discussed above, a general case for a sizable error is that some streaks fall close to or on top of galaxies or other objects, which can interfere with the streak fitting.

The coordinate error is defined as the difference between the ground-truth coordinates for the middle point of the streak and the corresponding coordinates given by \texttt{StreakDet}. For the brightest streaks, between magnitudes 20 and 21, the average coordinate error is 0.58 pixels, and the median coordinate error is 0.10 pixels, corresponding to 58 and 10 milliarcseconds, respectively. The coordinate error increases for both faint and long streaks. For all magnitudes, the average coordinate error is 1.92 pixels (192$\,{\rm mas}$), and the median coordinate error is 0.19 pixels (19$\,{\rm mas}$).

The primary source of the coordinate error is the length error. When \texttt{StreakDet} detects the length of the streak inaccurately, the middle point of the streak is also detected inaccurately. When taking only streaks whose length error is less than 0.1 pixels into account (22\% of all detections), the average coordinate error for all magnitudes decreases to 0.15 pixels (15$\,{\rm mas}$), and the median error decreases to 0.10 pixels (10$\,{\rm mas}$).

We tested \texttt{StreakDet} without astrometry and photometry at this point, as it was originally programmed to use UCAC4, and Gaia data are available for \textit{Euclid}. Therefore, there are no estimates for apparent magnitude errors.

\subsection{Multi-streak analysis}

\begin{figure*}%[h]
    \centering
    \includegraphics[width=\textwidth]{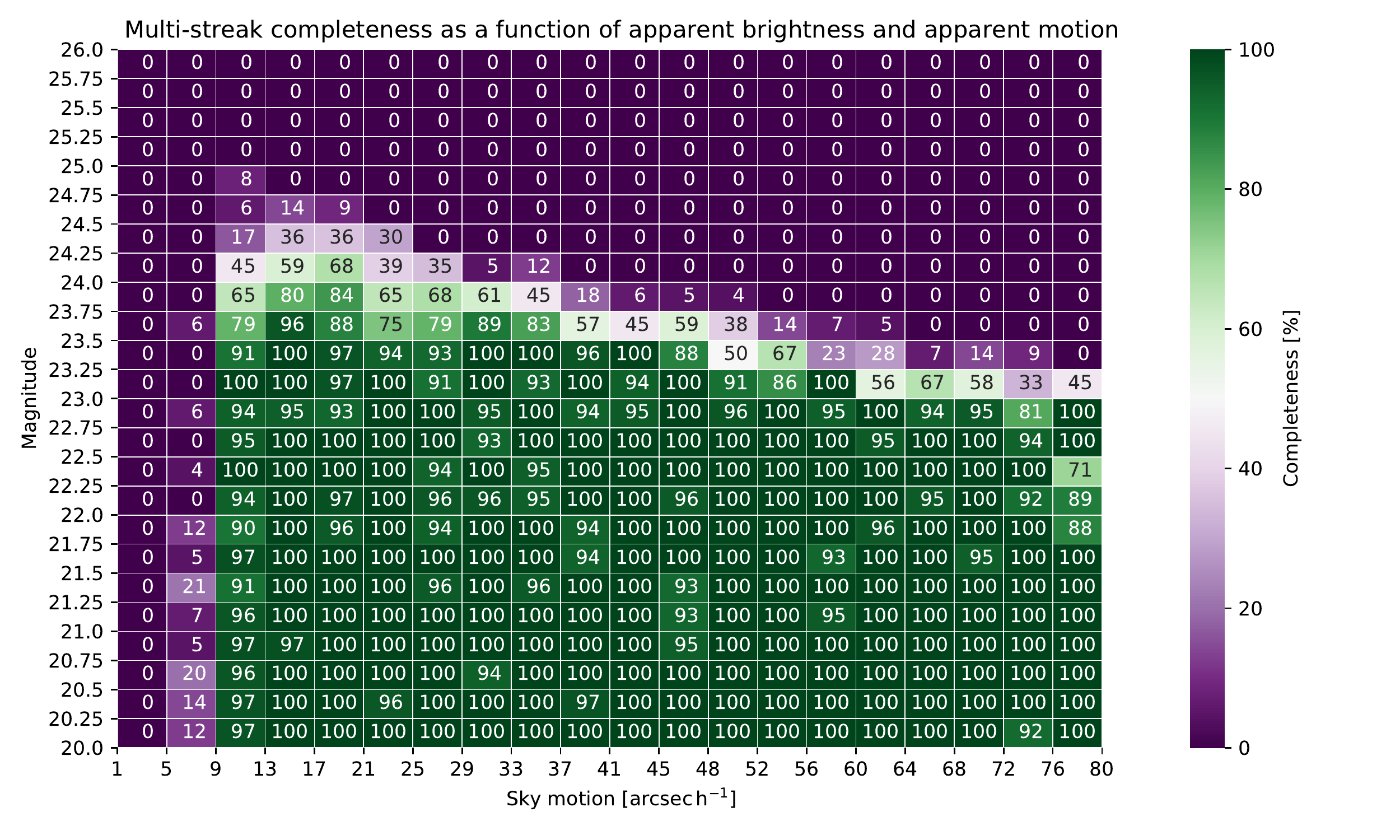}
    \caption{Multi-streak detection percentage as functions of apparent magnitude and apparent motion. The values on both axes show the limits of the bins. On average, there are 21 ground-truth multi-streaks per bin. The completeness shown is defined as the number of true-positive linked multi-streaks divided by the number of ground-truth multi-streaks consisting of at least two linked streaks. In other words, ground-truth streaks appearing in only one dither are ignored. If they were included, all the completeness values shown would be approximately 9\% lower.}
    \label{fig:Heatmap_multistreak}
\end{figure*}

Since the settings were optimized to maximize the detection percentage, when analyzing just the individual streaks, the number of false positives is high, and there are approximately 30 false positives for each true positive. Multi-streak analysis manages to discard most of the false positives, while maintaining most of the true positives. The multi-streak analysis results are shown in Fig. \ref{fig:Heatmap_multistreak} and in Tables \ref{table:mstable1} and \ref{table:mstable2}. The completeness is defined here as the fraction of multi-streaks found among the ground-truth multi-streaks that have visible streaks in at least two dithers. In other words, ground-truth objects that are visible in only one dither are ignored. These objects correspond to approximately 9\% of the simulated objects. If they were included, all the completeness values shown in the heat map would be approximately 9\% lower.

The parameters of multi-streak analysis are the following, with values we used in the analysis in the parentheses: minimum number of streaks in a multi-streak (2), maximum length difference between streaks (45\%), maximum angle difference between streaks (3\textdegree{}), maximum angle difference between the single streaks and the common multi-streak line (3\textdegree{}), maximum flux difference between streaks (29\%), and maximum difference in PSF sigma values (44\%). This set of parameters was found to retain virtually all of the true positives, while decreasing the number of false positives by three orders of magnitude to approximately 4\% of all \texttt{StreakDet} finds. The false-positive multi-streaks are typically caused by two individual false-positive streaks that happen to be approximately along the same line in two dithers and otherwise fulfill the filtering parameters.

The number of false positives can be decreased further by using stricter parameters, but the number of true positives starts to decrease faster than the number of false positives below a 4\% false-positive level. The most straightforward way to achieve virtually zero false positives is to require at least three linked streaks instead of two. This appears to guarantee 100\% purity, but the downside is a fairly large drop in the number of true positives, from 55.66\% to 44.36\% completeness in a full test set containing all tested streak lengths and magnitudes. The multi-streak analysis is a separate post-processing step, so it can be run multiple times with different settings, without the need to run \texttt{StreakDet} again. This way, multiple SSO data sets can be offered, for example, one with zero false positives but smaller completeness, and another with a small number of false positives but higher completeness. Currently, the multi-streak analysis takes approximately one minute to run for one pointing, 144 CCDs. With more optimization, the running time can be further reduced, if needed.

When comparing the top right and bottom right images in Fig. \ref{fig:Steps}, it seems that a substantial portion of the false positives are results of the streaks caused by the CTI shadows of the removed cosmic rays. As the OU-VIS data reduction pipeline considerably reduces the CTI effects in the images, the number of false-positive streaks should also decrease, increasing purity.

The fifth VIS exposure, which was not simulated in our data, can also help increase purity and completeness. However, the detectability of SSOs in the additional VIS exposure with \texttt{StreakDet} is limited to objects moving faster than $\sim60\,{\rm arcsec\,h^{-1}}$ because the exposure time of the additional VIS frame is only 116 seconds, instead of 565 s, resulting in much shorter streaks.

\begin{table} 
\caption{Multi-streak analysis results, combined for all tested data sets. Semi-hits refer to linked multi-streaks that contain both true and false-positive streaks. The table includes all the data underlying Fig. \ref{fig:Heatmap_multistreak}, including all the zero-bins, which explains the total completeness of only 55.66\%.}
\label{table:mstable1}
\centering
\begin{tabular}{lcc}
\hline\hline
Type &Number &Percentage \\
\hline
Ground-truth multi-streaks: &9737 & \\
\hline
\texttt{StreakDet} multi-streak finds: &5682 \\
\hline
Hits: &5420 &55.66\% \\
Semi-hits : &18 &0.18\% \\
Total hits: &5438 &55.85\% \\
Duplicates: &17 &0.17\% \\
Misses: &4299 &44.15\% \\
False positives: &227 &4.00\% of SD finds \\
\hline
\end{tabular}
\end{table}

\begin{table}
\caption{Detailed detection percentages of multi-streak analysis. GT refers to ground truth and SD refers to \texttt{StreakDet}.}
\label{table:mstable2}
\centering
\begin{tabular}{lcc}
\hline\hline
GT multi-streaks with 4 parts: &4009 &41.17\% of GTs \\
\hline
\quad Of which \texttt{StreakDet} found &2285 &57.00\% \\
\hline
\qquad With 4 SD hits: &1612 &40.21\% \\
\qquad With 3 SD hits: &491 &12.25\% \\
\qquad With 2 SD hits: &182 &4.54\% \\
\hline
GT multi-streaks with 3 parts: &4828 &49.58\% of GTs \\
\hline
\quad Of which \texttt{StreakDet} found &2692 &55.76\% \\
\hline
\qquad With 3 SD hits: &2211 &45.80\% \\
\qquad With 2 SD hits: &481 &9.96\% \\
\hline
GT multi-streaks with 2 parts: &900 &9.24\% of GTs \\
\hline
\quad Of which \texttt{StreakDet} found &443 &49.22\% \\
\hline
\qquad With 2 SD hits: &443 &49.22\% \\
\hline
\end{tabular}
\end{table}

\section{Conclusions and future work}

In the simulated \textit{Euclid} data, \texttt{StreakDet} finds a streak in at least one dither for 99.1\% of SSOs brighter than 23 magnitudes, with streak lengths longer than 15 pixels (10$\,{\rm arcsec\,h^{-1}}$), and it finds 96.9\% of all the individual streaks caused by these objects. For streaks fainter than magnitude 23, the finding percentage starts to decrease and reaches zero at magnitude 25. This is to be expected, given the limiting magnitudes of VIS. Still, there appears to be some room for improvement in the detection percentage above magnitude 23 since \texttt{StreakDet} fails to detect some faint streaks that are still visible to the naked eye, and it finds some streaks of similar brightness, but not others. For streaks shorter than 13 pixels (8$\,{\rm arcsec\,h^{-1}}$), the detection percentage declines rapidly, approaching zero at streaks lengths below 9 pixels (6$\,{\rm arcsec\,h^{-1}}$). The multi-streak analysis run on the final \texttt{StreakDet} results worked well in retaining most of the true positives, while removing most or, with stricter settings, all of the false positives.

Going forward, the areas with the greatest potential for improving the overall detection ability of asteroids in \textit{Euclid} data are improving the segmentation phase of \texttt{StreakDet}, making it more readily able to detect short and/or faint streaks. Improving the detection percentage for faint streaks is especially important, because most of the asteroids appearing in \textit{Euclid} images are close to the limiting magnitude. The simulated data used in this work contained too few faint galaxies, compared to the real sky. This could somewhat change the expected \texttt{StreakDet} detection percentage for streaks above magnitude 24. However, as \texttt{StreakDet} currently detects very few of the streaks that are that faint, the effect on results should be small, and improving the detection of faint streaks with realistic faint galaxy density remains for future work in any case. If possible, it would also be useful to optimize the \texttt{StreakDet} code so that is runs more rapidly, because currently, it takes approximately one minute to analyze one 4\,k\,$\times$\,4\,k image. Another option is to develop an advanced deep learning model, such as a convolutional neural network, that is capable of directly returning the coordinates of the asteroids in the images. One possible option is to combine these two approaches. From a practical point of view, it is probably necessary to collect at least some non-simulated \textit{Euclid} training data for a machine-learning approach using a method such as \texttt{StreakDet}. Another prospect is studying the analysis of asteroids in the NISP images, in terms of detection and measurement of the spectral energy distributions.

\begin{acknowledgements}

We thank the anonymous referee for constructive comments.

M.P. and M.G. acknowledge funding from the Academy of Finland (projects \#316292 and \#299543).

The Euclid Consortium acknowledges the European Space Agency and the support of a number of agencies and institutes that have supported the development of \textit{Euclid}. A detailed complete list is available on the \textit{Euclid} web site (\texttt{http://www.euclid-ec.org}). In particular, the Academy of Finland, the Agenzia Spaziale Italiana, the Belgian Science Policy, the Canadian Euclid Consortium, the Centre National d'Etudes Spatiales, the Deutsches Zentrum f\"ur Luft- und Raumfahrt, the Danish Space Research Institute, the Funda\c{c}\~{a}o para a Ci\^{e}ncia e a Tecnologia, the Ministerio de Economia y Competitividad, the National Aeronautics and Space Administration, the Netherlandse Onderzoekschool Voor Astronomie, the Norwegian Space Agency, the Romanian Space Agency, the State Secretariat for Education, Research and Innovation (SERI) at the Swiss Space Office (SSO), and the United Kingdom Space Agency.

\end{acknowledgements}

%-------------------------------------------------------------------
\bibliographystyle{aa} % style aa.bst
\bibliography{SD_for_Euclid.bib}

%\begin{thebibliography}{}

%\end{thebibliography}

\end{document}